\documentclass[a4paper,fleqn,usenatbib]{mnras}
\usepackage{amssymb,amsmath,graphicx,psfrag,mathtools}
\usepackage[T1]{fontenc} %MNRAS
\usepackage{ae,aecompl} %MNRAS
\usepackage{url}
\usepackage{revsymb}
\hypersetup{draft} %TO AVOID ERROR: \PDFENDLINK ENDED UP IN DIFFERENT
\title[Radiation Modes in FRB 20220912A Microshots]{Radiation Modes in FRB 20220912A Microshots and a Crab PSR Nanoshot}
\author[J. I. Katz]{
J. I. Katz,$^{1}$\thanks{E-mail katz@wuphys.wustl.edu} %MNRAS
\\
$^{1}$Department of Physics and McDonnell Center for the Space Sciences,
Washington University, St. Louis, Mo. 63130 USA %MNRAS
}
\date{Accepted XXX.  Received YYY; in original form ZZZ} %MNRAS
\pubyear{2020} %MNRAS
\date{\today}
\begin{document} %MNRAS
%\psfrag{theta}{$\theta$}  Works only on COMPLETE strings
\label{firstpage} %MNRAS
\pagerange{\pageref{firstpage}--\pageref{lastpage}} %MNRAS
\maketitle %MNRAS
\begin{abstract}
	A microshot from FRB 20220912A \citep{H23} satisfies the uncertainty
	relation $\Delta \omega \Delta t \ge 1$ by a factor of only 
	$\lessapprox 3$.  A Crab pulsar nanoshot \citep{HE07} exceeds this
	bound by a similar factor.  The number of orthogonal plasma modes
	contributing to the coherent radiation is also $\approx \Delta
	\omega \Delta t$, placing constraints on their excitation and growth.
\end{abstract}
 %MNRAS
\section{Introduction}
The recent discovery in Westerbork observations of FRB 20220912A \citep{H23}
of microshots with temporal width $\le 31.25\,$ns in a spectral channel of
width $16\,$MHz leads to the question of how close microshots can approach
the uncertainty bound $\Delta \omega \Delta t = 2 \pi \Delta \nu \Delta t
\gtrsim 1$.  The instrumental temporal and spectral resolutions are only
upper bounds on the actual pulse duration and bandwidth, so only an upper
bound $\Delta \omega \Delta t \lesssim 3$ can be set.

The Crab pulsar produces occasional ``nanoshots'' \citep{HE07}.  In the most
extreme example (their Fig. 5) the temporal width $\le 0.2\,$ns, and $\Delta
\nu \le 2.2\,$GHz.   Then $\Delta \omega \Delta t \lesssim 3$, similar to
the bound inferred for the FRB microshots.

This uncertainty bound is inescapable mathematics, although its quantitative
value depends on the pulse shape and the definitions of $\Delta \omega$ and
$\Delta t$.  For a Gaussian pulse, if $\Delta \omega$ and $\Delta t$ are
defined as full widths at $1/e$ of maximum, then
\begin{equation}
	\label{uncert}
	\Delta \omega \Delta t = {4 \over \pi}.
\end{equation}
The condition
\begin{equation}
\Delta\omega\Delta t = 1
\end{equation}
is satisfied for a Gaussian pulse if $\Delta \omega$ and $\Delta t$ are
defined as the full widths at $\exp{-\sqrt{\pi/4}} \approx 0.412$ of
maximum.

If $\Delta \omega \Delta t > 1$ we consider the natural generalizations of
the Gaussian to larger widths, the Gauss-Hermite functions, the eigenstates
of the one-dimensional harmonic oscillator with Hamiltonian
$H = (kx^2/2+p^2/2m)$:
\begin{equation}
	G_n(x) = H_n(x) \exp{(-x^2/2)},
\end{equation}
where $H_n(x)$ is the $n$-th Hermite polynomial.  Like the Gaussian
($G_0(x)$), the Gauss-Hermite functions are their own Fourier transforms
\citep{C92,H06}, as they must be because the harmonic oscillator Hamiltonian
is symmetric under the interchange $kx^2 \longleftrightarrow p^2/m$.  Here
we replace $\sqrt{k}x$ by $t$ and $p/\sqrt{m}$ by $\omega$.

It is readily seen, either from the properties of the Hermite polynomials
or from the fact that the $n$-th excited state of the harmonic oscillator
has energy $E_n = (n + 1/2) \hbar \omega$ and classical width 
$\sqrt{2E_n/k}$, that for these more complex pulse frequency and temporal
profiles
\begin{equation}
	\label{n}
	\Delta \omega \Delta t \sim n.
\end{equation}
Eq.~\ref{n} gives the approximate number of orthogonal eigenmodes whose
superposition makes a pulse with $\Delta \omega \Delta t > 1$.
\section{The Observations}
The lower panel of Fig.~5 of \citet{H23} shows microshots in the 1304 MHz
band at approximately $11\,\mu$s (with respect to the arbitrary zero time
of the plot) whose temporal widths are no more than a single 31.25 ns
resolution element and that are almost completely confined to a single 16
MHz wide spectral band.  Their intensity in the 1288 MHz band is at least an
order of magnitude less than in the 1304 MHz band and their intensity in the
1320 MHz band is a few times less than at 1304 MHz, implying an intrinsic
$\Delta \nu$ less than the band spacing of 16 MHz (the quantitative value
depending on the assumed spectral shape).  Conservatively taking $\Delta \nu
= 16\,$MHz implies $\Delta \omega \Delta t \lessapprox 3$.  By Eq.~\ref{n},
no more than about three modes of the electromagnetic field, and of the
plasma waves that coherently radiated it, contributed significantly to the
observed microshot.  A similar result holds for a $\le 0.2\,$ns nanoshot of
the Crab pulsar \citep{HE07}.
\section{Plasma Physics}
Coherent emission, necessary to explain the extraordinary brightness
temperatures of FRB \citep{K14} and of pulsars, requires ``bunching'' of
the radiating charges that must result from the exponential growth of plasma
waves.  The many $e$-foldings of exponential growth raise the the amplitudes
of the few fastest growing modes far above those of other modes.  An
observed brightness temperature of $10^{36}\,$K requires $N \gtrapprox
50\,e$-foldings if the initial brightness temperature was $m_e c^2/k_B$
(an arbitrary but plausible initial thermal value).
\subsection{Linear Growth}
Assume a plasma instability grows exponentially with growth rate
\begin{equation}
	\gamma(\zeta) \sim \gamma_0 \left(1 - {(\zeta-\zeta_{max})^2
	\over (\Delta \zeta)^2}\right),
\end{equation}
where $\zeta$ is a parameter of the plasma wave (perhaps its wave-vector)
and $\Delta \zeta$ the width of $\gamma(\zeta)$.  This is a general form
that assumes nothing about the specifics of the plasma instability, but
that does require that the plasma modes interact only weakly; the
governing equations can be linearized so that eigenmodes grow exponentially
and essentially independently of each other.  From the fact that after $N$
$e$-folds the width of the microshots of FRB 20220912A
\begin{equation}
	\label{width1}
	{|\zeta-\zeta_{max}| \over \Delta \zeta} \approx \sqrt{1/N} \approx
	0.15,
\end{equation}
the fact that $\Delta \nu/\nu \approx 0.01$, and the plausible assumption
that $d\zeta/d\nu \sim \zeta/\nu$,
we estimate
\begin{equation}
	\label{width2}
	{\Delta \zeta \over \zeta_{max}} \sim \sqrt{N}
	{|\zeta-\zeta_{max}| \over \zeta_{max}} \sim 0.07.
\end{equation}
This is an approximate constraint that can be placed on any linearized
theory of the plasma instability.
\subsection{Coupled Waves}
In an alternative model, the plasma waves are strongly coupled as they
approach saturation, resembling a soliton \citep{ZK65}, rather than being
described as a superposition of several weakly interacting eigenmodes.  This
would suggest that $\Delta\omega\Delta t \approx 1$ because the radiation is
produced by a single nonlinear wave, rather than by the sum of multiple
weakly interacting waves; $n \approx 1$ would be required.  The observations
of FRB microshots \citep{H23} and Crab pulsar nanoshots \citep{HE07} cited
here are consistent with this hypothesis, which is supported by the fact
that both of these extreme phenomena are described by similar bounds on
$\Delta\omega\Delta t$.
\section{Base-Band Signals}
If base-band voltages are measured in a microshot or nanoshot, they will
show $\sim \omega \Delta t \sim \Delta\omega\Delta t(\omega/\Delta\omega)
\sim n(\omega/\Delta\omega)$ cycles of oscillation.  This is a mathematical
consequence of $\Delta \omega \Delta t \sim n$, and does not depend on a
physical model.  However, such a direct measurement might provide other
illuminating information about the radiation process and the plasma physics
that drives it.  The dependence of electric field on time may permit
distinguishing the weakly and strongly coupled models.

\begin{figure}
	\centering
	\includegraphics[width=0.99\columnwidth]{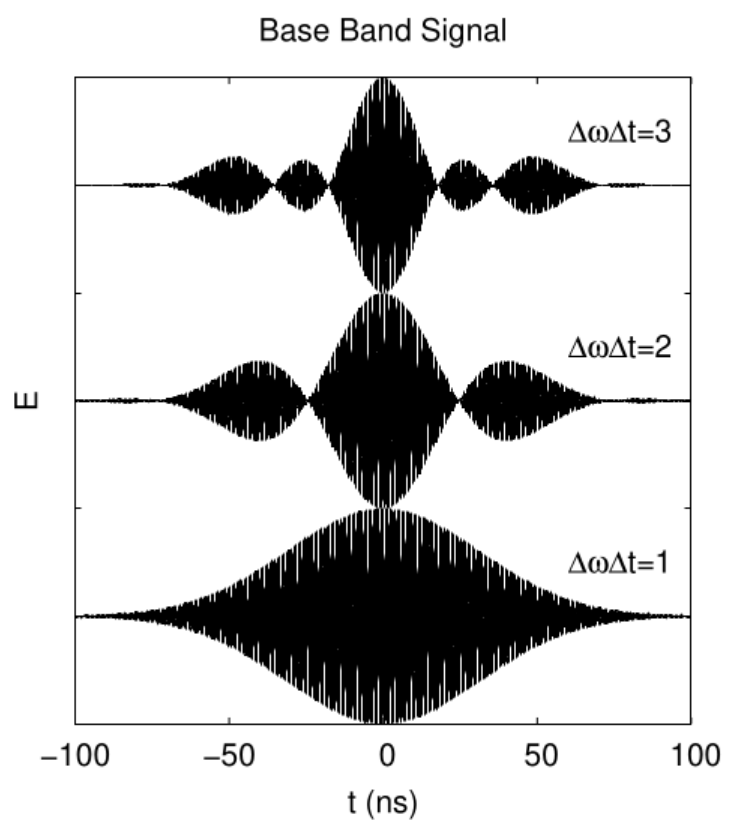}
	\caption{\label{base}Baseband signals for $\Delta\omega\Delta t = n
	= 1, 2, 3$ and $\Delta t = 31.25\,$ns.  The wider bandwidth pulses
	have narrower main peaks (a mathematical necessity).  The secondary
	peaks may be observable as frequency-averaged functions of intensity
	{\it vs.\/} time even if noise prevents measurement of the baseband
	signal.  Fine structure is the result of beats between the signal
	and the discrete resolution of the graphics screen, and is not
	physical.}
\end{figure}

As an example of what might be seen in base-band signals, Fig.~\ref{base}
shows the time dependence of the electric field for signals with three values
of $n = \Delta\omega\Delta t$, where $\Delta t$ is taken as the 31.25 ns
temporal resolution of the data reported by \citet{H23}.  For $n=1$ we show
a cosine with nominal $\nu_0 = 1308\,$MHz; for $n=2$ we take the mean of two
equally weighted cosines at $\nu = (1308 \pm 10.2)\,$MHz, with
r.m.s.~$\Delta\omega = 2/\Delta t$; for $n=3$ we take the mean of three
equally weighted cosines at $\nu = [1308 + (-1,0,1) \times 18.7]\,$MHz, with
r.m.s.~$\Delta\omega = 3/\Delta t$.  Each of these profiles is multiplied by
an envelope function $\exp{[-(t^2/2(\Delta t)^2)]}$.

Even if the signal-to-noise ratio is insufficient to resolve the base-band
oscillations, their envelope may reveal the number of contributing modes
$n = \Delta\omega\Delta t$.  The signal-to-noise ratio of the envelope is
greater than that of the base-band signal by a factor ${\cal O}
(\sqrt{\omega_0/n\Delta\omega})$, which is ${\cal O}(16/\sqrt{n})$ for the
parameters shown; use of a matched filter would eliminate the factor of
$1/\sqrt{n}$.
\section{Discussion}
In a weakly coupled wave model, because $N$ is large a narrowly peaked
distribution of amplitudes of modes with $\Delta \nu \ll \nu$ does not
require a narrowly peaked growth rate.  Eq.~\ref{width1} may be inverted to
find $N$ from the spectral width of a FRB microshot; spectral narrowness
demands many $e$-folds of exponential growth of the underlying plasma
waves. 

The secondary intensity peaks (in the envelope of the $\approx 1308\,$MHz
oscillation in Fig.~\ref{base}) are required by values of $n = \Delta\omega
\Delta t \gtrsim 2$.  Hence the number $n$ of contributing modes may be
inferred from the time dependence of the intensity if it is sufficiently
resolved, as well as from direct measurement of $\Delta\omega$ and$\Delta t$
of the central peak.  The temporal resolutions of the microshots of
\citet{H23} and the nanoshot of \citet{HE07} were insufficient to determine
$n$, but future observations with better temporal resolution should be able
to show this structure and determine the number of contributing modes.
\section*{Acknowledgment}
I thank the late K. Case for long ago teaching me about functions that are
their own Fourier transforms.
\section*{Data Availability}
This theoretical study did not generate any new data.

\label{lastpage} %MNRAS

\begin{thebibliography}{99}
	\bibitem[\protect\citeauthoryear{Cincotti, Gori \& Santasiero}
		{1992}]{C92} Cincotti, G., Gori, F. \& Santasiero, M. 1992
		{\it J. Phys. A\/} 25, L1191.
	\bibitem[\protect\citeauthoryear{Hankins \& Eilek}{2007}]{HE07}
		Hankins, T. H. \& Eilek, J. A. 2007 \apj\ 670, 693.
	\bibitem[\protect\citeauthoryear{Hewitt {\it et al.\/}}{2023}]{H23}
		Hewitt, D. M., Hessels, J. W. T., Ould-Boukattine, O. S.
		{\it et al.\/} 2023 arXiv:2308.12118.
	\bibitem[\protect\citeauthoryear{Horikis \& McCallum}{2006}]{H06}
		Horikis, T. P. \& McCallum, M. S. 2006 {\it J. Opt. Soc.
		Am. A\/} 23, 829.
	\bibitem[\protect\citeauthoryear{Katz}{2014}]{K14} Katz, J. I. 2014
		\prd\ 89, 103009.
	\bibitem[\protect\citeauthoryear{Zabusky \& Kruskal}{1965}]{ZK65}
		Zabusky, N. J. \& Kruskal, M. D. 1965 \prl\ 15, 240.
\end{thebibliography}
\end{document}